# HIGH ENERGY PARTICLE COLLIDERS: PAST 20 YEARS, NEXT 20 YEARS AND BEYOND

VLADIMIR D. SHILTSEV

*Fermilab Accelerator Physics Center, PO Box 500, MS221, Batavia, IL, 60510, USA*

*Abstract:* Particle colliders for high energy physics have been in the forefront of scientific discoveries for more than half a century. The accelerator technology of the collider has progressed immensely, while the beam energy, luminosity, facility size and the cost have grown by several orders of magnitude. The method of colliding beams has not fully exhausted its potential but its pace of progress has greatly slowed down. In this paper we very briefly review the method and the history of colliders, discuss in detail the developments over the past two decades and the directions of the R&D toward near future colliders which are currently being explored. Finally, we make an attempt to look beyond the current horizon and outline the changes in the paradigm required for the next breakthroughs.

<u>Content:</u>

# Chapter 1: Introduction, Colliders of Today

Particle accelerators have been widely used for physics research since the early 20th century and have greatly progressed both scientifically and technologically since then. To gain an insight into the physics of elementary particles, one accelerates them to very high kinetic energy, let them impact on other particles, and detect products of the reactions that transform the particles into other particles. It is estimated that in the post-1938 era, accelerator science has influenced almost 1/3 of physicists and physics studies and on average contributed to physics Nobel Prize-winning research every 2.9 years [1]. Colliding beam facilities which produce high-energy collisions (interactions) between particles of approximately oppositely directed beams did pave the way for progress since the 1960's.

The center of mass (CM) energy $E_{cm}$ for a head-on collision of two particles with masses $m_1$, $m_2$ and energies $E_1$ and $E_2$ is

$$E_{cm} = \left[2E_1E_2 + (m_1^2 + m_2^2)c^4 + 2\sqrt{E_1^2 - m_1^2c^4}\,\sqrt{E_2^2 - m_2^2c^4}\right]^{1/2} . \qquad (1)$$

For many decades, the only arrangement of accelerator experiments was a fixed target setup where a beam of particles accelerated with a particle accelerator hit a stationary target set into the path of the beam. In this case, as follows from Eq. (1), for high energy accelerators $E >> mc^2$, the CM energy is $E_{cm} \approx (2Emc^2)^{1/2}$. For example, $E$=1000 GeV protons hitting stationary protons $mc^2 \approx 1\ GeV$ can produce reactions with about 43 GeV energy. A more effective colliding beam set-up in which two beams of particles are accelerated and directed against each other, has much higher center of mass energy of $E_{cm} \approx 2(E_1E_2)^{1/2}$. In the case of two equal masses of particles (e.g. protons and protons, or protons and antiprotons) colliding with the same energy $E$ of 1000 GeV, one gets $E_{cm}=2E$ or 2000 GeV. Such an obvious advantage led to the first practical proposals of colliding-beam storage rings in the late 1950's [2,3].

Almost three dozen of colliders reached operational stage between the late 50's and now. Schematic drawings of several collider types are shown in Fig.1. In storage ring configurations - Fig.1a and 1b – particles of each beam circulate and repeatedly collide. This can be done in a single ring if the beams consist of the same energy antiparticles. In linear colliders, first proposed in Ref. [4], the beams are accelerated in linear accelerators (linacs) and transported to a collision point: either in the simple two linac configurations depicted in Fig.1c, or with use of the same linac and two arcs as in Fig.1d. Another possible linac-ring configuration is shown in Fig.1e.

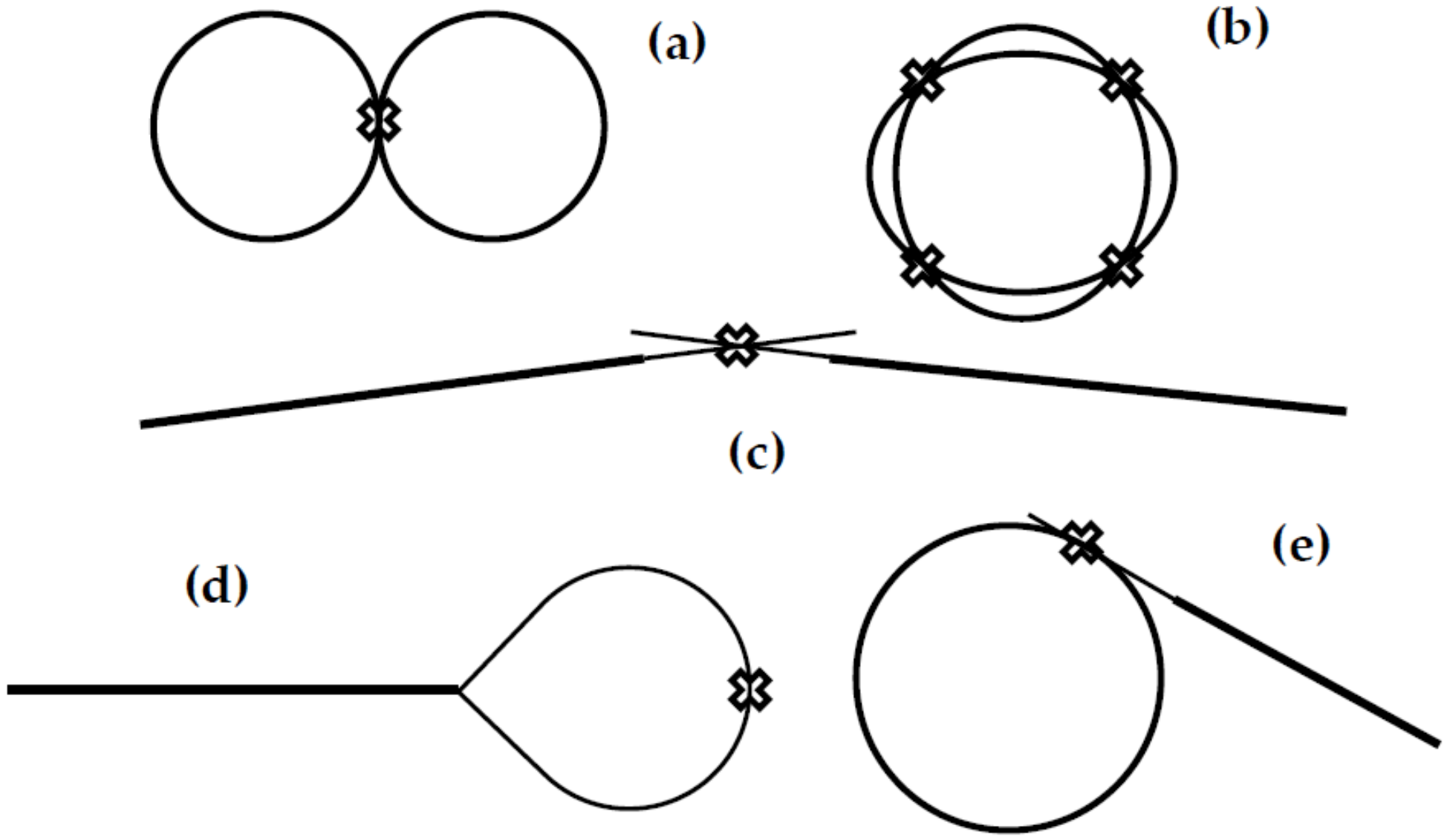

Figure 1: Schematics of particle collider types.

The first colliding lepton beam facilities were built in the early 1960s almost simultaneously at three laboratories: *e-e-* colliders AdA at the Frascati laboratory near Rome in Italy, the VEP-1 collider in the Novosibirsk Institute of Nuclear Physics (USSR) and the Princeton-Stanford Colliding Beam Experiment at Stanford (USA). Their center of mass energies were 1 GeV or less. Construction of the first hadron (proton-proton) collider, the Intersecting Storage Rings, began at CERN (Switzerland) in 1966, and in 1971, this collider was operational and eventually reached $E_{cm}$=63 GeV. The first linear collider was the *e-e+* SLAC Linear Collider (SLC) constructed at Stanford in the late 1980's. Detail discussions on the history of colliders can be found, e.g., in [5, 6], Ref.[7] gives an overview of the development of the colliding beams facilities in Novosibirsk. (As this article is not intended to be a historical overview, only the most recent relevant references will be provided below.)

The energy of colliders has been increasing over the years as demonstrated in Fig.2. There, the triangles represent maximum CM energy and the start of operation for lepton (usually, *e+e-* ) colliders and full circles are for hadron (protons, antiprotons, ions, proton-electron) colliders. One can see that until the early 1990's, the CM energy on average increased by a factor of 10 every decade and, notably, the hadron colliders were 10-20 times more powerful. Since then, following the demands of high energy physics, the paths of the colliders diverged to reach record high energies in the particle reaction. The Large Hadron Colider (LHC) was built at CERN, while new *e+e-* colliders called “particle factories” were focused on detail exploration of phenomena at much lower energies.

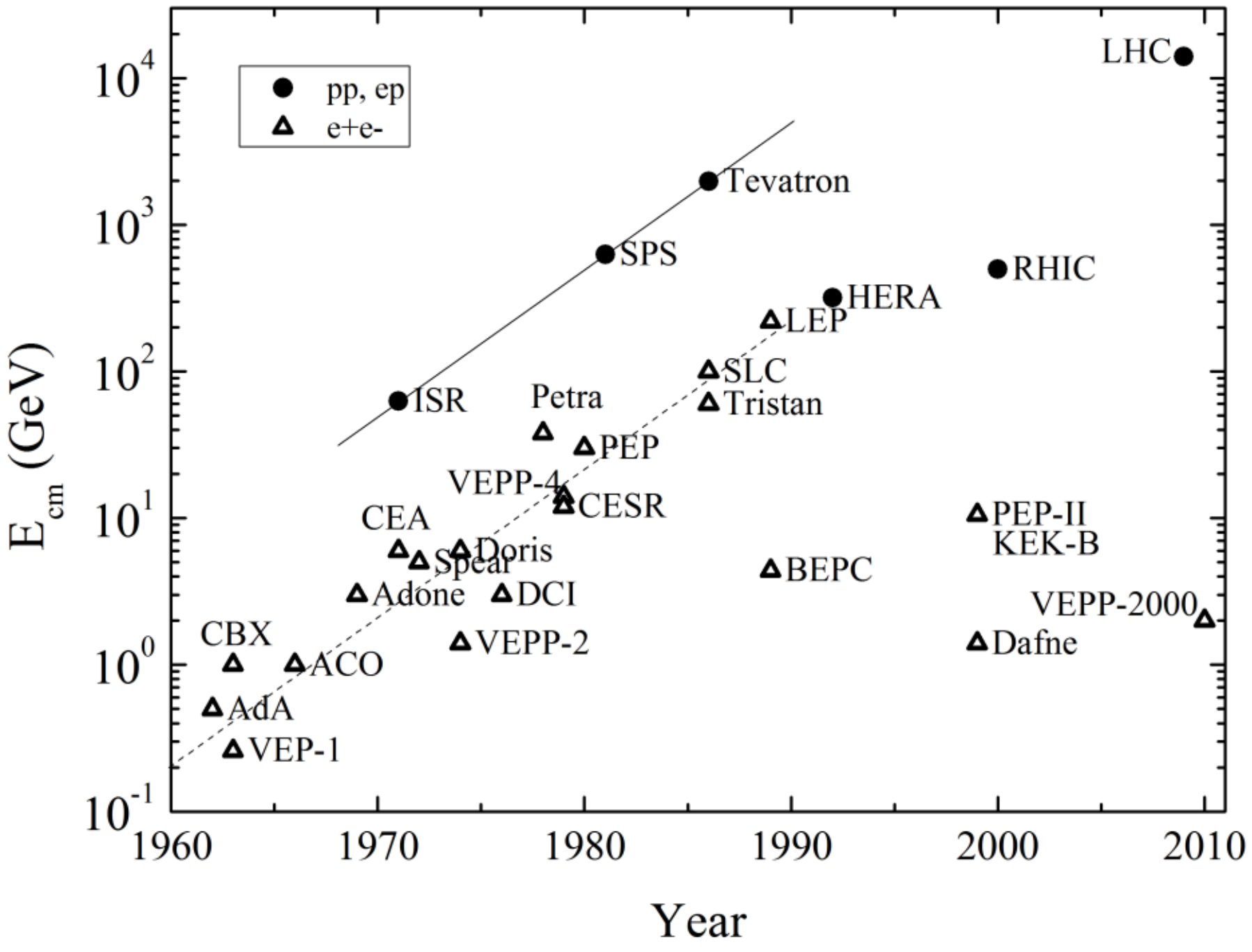

Figure 2: Colliders over the decades.

The exploration of rare particle physics events require appropriately high energy but also sufficiently high number of them. The event rate $dN_{\mathrm{exp}}/dt$ in a collider is proportional to the interactions cross-section $\sigma_{\mathrm{int}}$ and the factor of proportionality is called the luminosity:

$$\frac{dN_{\mathrm{exp}}}{dt} = L \cdot \sigma_{\mathrm{int}} \ , \qquad (2)$$

If two bunches containing $N_1$ and $N_2$ particles collide with frequency $f$, the luminosity is:

$$L = f\,\frac{N_1 N_2}{A} \quad , \qquad (3)$$

where $A$ is an effective overlap area of the beams. In the simplest case of two bunches with identical Gaussian transverse beam profiles characterized by rms widths of $\sigma_x$ and $\sigma_y$, the overlap area is approximately equal to $A=4\pi\sigma_x\sigma_y$ (we omit here any corrections due to non-uniform longitudinal profile of the luminous region.) The beam size can in turn be expressed in terms of the rms normalized transverse emittance $\varepsilon$ (which is an approximate adiabatic invariant of particle motion during acceleration) and the *amplitude function* $\beta$ (which is a beam optics quantity determined by accelerator transverse, most often magnetic, focusing system):

$$\sigma^2 = \frac{\varepsilon}{\gamma\beta} \quad , \tag{4}$$

$\gamma = E/mc^2$ is the relativistic Lorentz factor. So, the basic equation for the luminosity (3) can now be re-written in terms of emittances and the amplitude functions at the interaction point (which we denote by asterisks) as

$$L = f\gamma \frac{N_1 N_2}{4\pi\sqrt{\varepsilon_x \beta_x^* \varepsilon_y \beta_y^*}} \quad . \tag{5}$$

Therefore to achieve high luminosity, one has to maximize population of bunches with as low as possible emittances and to collide them at high frequency at locations where the focusing beam optics provide the lowest values of the amplitude functions $\beta^*_{x,y}$. Increasing the beam energy and thus, factor $\gamma$ in Eq.(5), is, generally speaking, of help, too.

Figure 3 demonstrates impressive progress of luminosities of colliding beam facilities since the invention of the method Again, the triangles are lepton colliders and full circles are for hadron colliders. One can see that over the last 50 years, the performance of the colliders has improved by more than 6 orders of magnitude and reached record high values of over $10^{34}$cm$^{-2}$s$^{-1}$. At such luminosity, one can expect to produce, e.g., 100 events over one year of operation (about $10^7$ s) if the reaction cross section is 1 femtobarn (fb)=$10^{-39}$ cm$^2$.

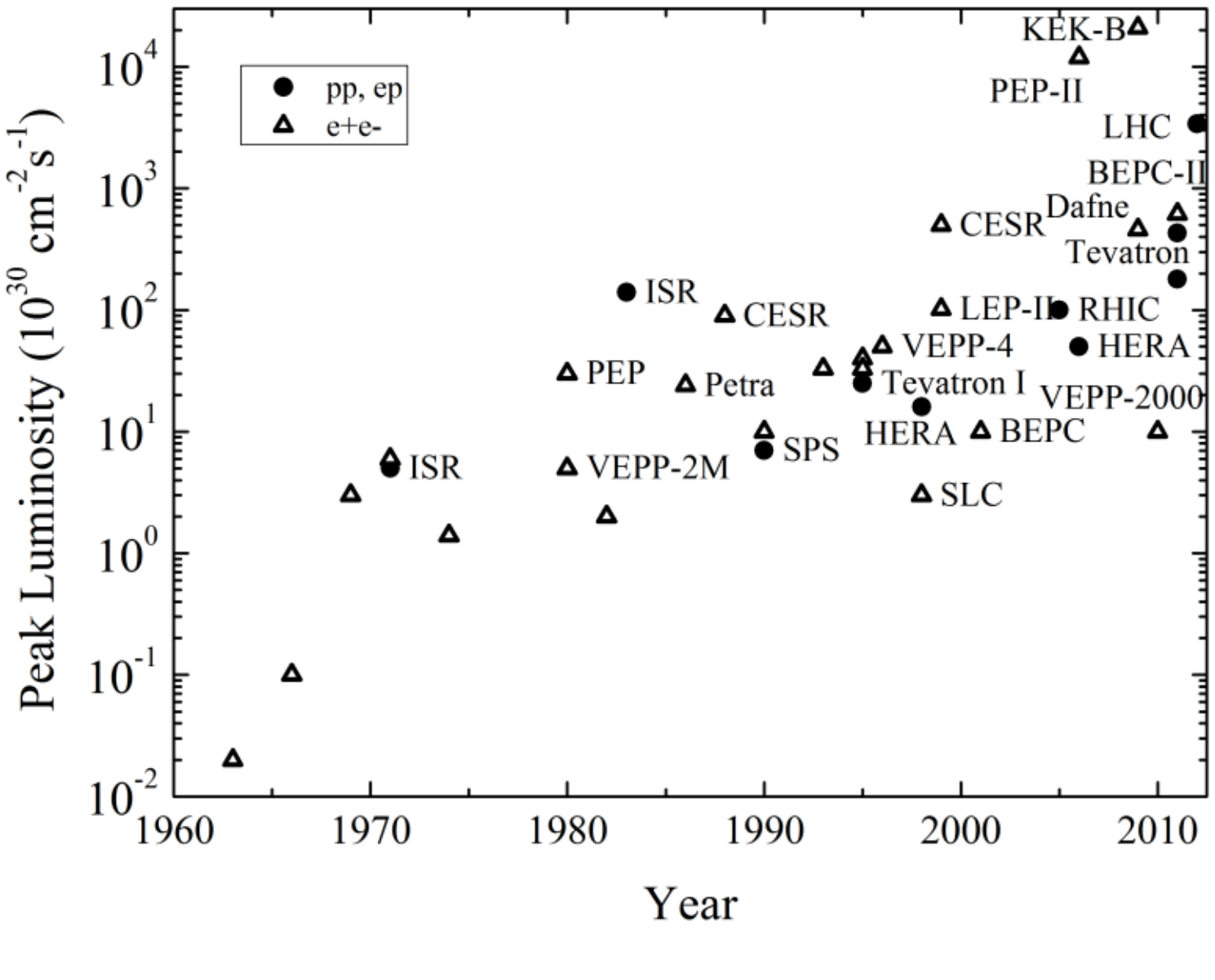

Figure 3: Peak luminosities of particle colliders.

Needless to say such great progress in both the energy and the luminosity of colliders has come from numerous advances in accelerator science and technology. The format of this article does not allow us to go through all the advances in detail – an interested reader can be referred for details to, e.g., a comprehensive handbook [6] – but it is worth at least a list major ones, as they will be relevant to our subsequent discussion on future colliders. References provided below will be given only to relatively recent development of the last two decades.

The maximum energy of colliders is determined by practical considerations, of which the first is the size of the facility. For a linear collider, beam energy is product of average accelerating gradient $G$ and length of the linac $l$:

$$E = eG \cdot l \quad . \tag{6}$$

In (the only) linear collider SLC, the average gradient in a 3-km-long linac, which was powered at a frequency of 2.856 GHz ($\lambda_{RF}$=10.5cm RF wavelength in normal-conducting copper structures), did eventually reach some 21 MV/m. Notably, as the length of the linac did not change, the beam energy approximately tripled over 30 years of operation because of upgrades, such as quadrupling the number of RF power sources (pulsed klystrons), increasing the peak output power of the klystrons from 35 MW to 65 MW, and further increasing the power by compressing the length of the RF pulses.

In circular colliders, the maximum momentum and energy of ultra-relativistic particle is determined by the radius of the ring $R$ and average magnetic field $B$ of bending magnets:

$$pc = eB \cdot R \quad \text{or} \quad E[GeV] = 0.3 \cdot B[T] \cdot R[m]. \tag{7}$$

Again, evolution of energy was driven by practical considerations: e.g., maximum field of normal conducting magnets of about 2T at some moment was not adequate for the energy demands because of required longer accelerator tunnels and increasing magnet power consumption. Development of superconducting (SC) magnets – see Fig.4 - which employ high electric current carrying NbTi wires cooled by liquid Helium below 5K, opened the way to higher fields and record high energy hadron colliders [8]. The latest of them, 14 TeV c.m. LHC at CERN uses 8.3T double bore magnets in 26.7 km circumference tunnel.

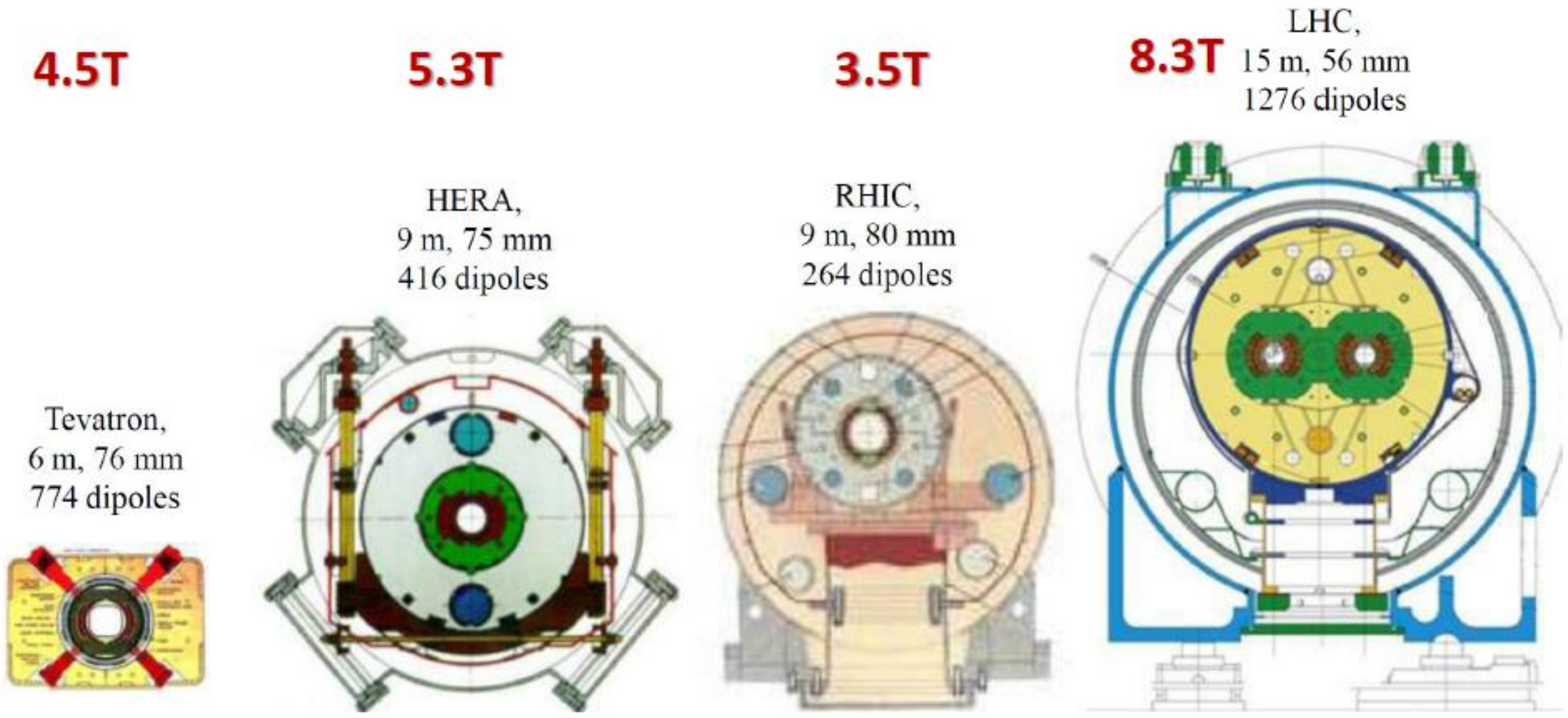

Figure 4: Superconducting dipole magnets for high energy hadron colliders: Tevatron (NbTi, warm-iron, small He plant, 4.5K), HERA (NbTi, Al collar, cold iron), RHIC (simple and economical design) and LHC (2K super fluid He, double bore). Courtesy A.Zlobin.

To remain superconducting, such magnets need to operate within very strict limits on the power deposited into the low-temperature components (vacuum pipes, cold iron, SC cable, etc) – typically on the order of 1 W/m or less, and that makes them of no practical use in high energy lepton accelerators, as relativistic electrons and positrons quickly lose energy due to the synchrotron radiation :

$$\delta E = \frac{4\pi}{3}\frac{e^2\gamma^4}{R} = 88.5[\frac{keV}{turn}]\cdot\frac{E^4[GeV]}{R[m]} \quad (8).$$

The total power radiated into the walls reaches becomes prohibitively high, and, e.g., 22 MW or about 800 W/m even in the largest radius *e+e-* collider LEP (in the same tunnel, which is now occupied by LHC), at the beam energy of 105 GeV and relatively low average beam current of 4mA. Besides the need to replenish electron beam power by accelerating RF cavities, the synchrotron radiation leads to significant heating and outgassing of the beam vacuum pipe. On the other hand, the attainment of sufficiently long lifetimes of continuously circulating beams requires gas pressures of 1-10 nTorr or better. This technological challenge, though, has been successfully resolved in modern (lower energy) colliding “factories” operating with multi-Ampere beams. Radiation of protons (ions) is smaller by a significant factor $(\gamma_p/\gamma_e)^4=(m_e/m_p)^4$ – see Eq.(8) - but still can be of a certain concern at very high energy, high current SC accelerators, e.g., LHC.

From the very beginning of the colliders, it was understood that their effective operation required stability of particle's motion in the rings. It is desired for particles to exhibit bound transverse oscillations in the fields of focusing magnets, e.g., as in the solution of Hill's equation

$$x(s) = A\sqrt{\beta(s)}\cos[\psi(s)+\delta] , \tag{9}$$

where the phase $\psi(s)$ advances around the ring and determines so called tune, or the number of such betatron oscillations per turn:

$$\nu = \frac{1}{2\pi}\oint d\psi(s) \quad . \tag{10}$$

Imperfections of the guiding magnetic fields can lead to multi-turn instabilities due to linear or non-linear resonances and cause particles to leave the acceptable aperture. At first, this was avoided by careful choice of the tune(s) – which is usually set away from integer or rational values $\nu \neq n/m$ - and with careful construction of the magnets. Magnetic field quality is given by the multipole coefficients in the expansion:

$$B_x + i\cdot B_y = B_0 \sum_{n=0} (b_n + ia_n)\left[\frac{x+iy}{R_0}\right]^n \quad , \tag{11}$$

where $R_0$ is the reference radius, the pole number is $2(n+1)$ and $b_n(a_n)$ are the normal (skew) multipole coefficients, and $b_0$ is unity. Over the years, collider magnet builders have perfected the magnet designs to the level of few $10^{-4}$ in the undesired multipoles – this was particularly challenging for SC magnets where such quality required very tight tolerances of less than a few dozen microns for the placements and position stability of the conductor coils under enormous magnetic forces [8]. Even more demanding were the field quality tolerances for the special, very strong magnets now routinely used in modern colliders to over-compress beams at the interaction points and arrange minimal possible beta-functions at the collision points $\beta_{x,y}^*$ - per Eq. (5) such magnets have greatly helped to achieve higher luminosities.

Unfortunately, the growing demand for luminosity led to orders of magnitude stronger non-linearity due to the very nature of the colliders – namely, due to electric and magnetic forces of the opposite bunch at the interaction points. The nonlinearity due to the collisions is characterized by dimensionless parameter called the “beam-beam parameter” or the “beam-beam tune shift”:

$$\xi_{x,y} = \frac{N r_c}{2\pi\gamma(\sigma_x^* + \sigma_y^*)} \cdot \frac{\beta_{x,y}^*}{\sigma_{x,y}^*} \quad , \tag{12}$$

(here, all the parameters are for the opposite bunch, and $r_c=e^2/mc^2$ is the classical radius of the particles) and can reach the value of $\xi$~0.03 in hadron colliders to ~0.13 in lepton colliders. Beyond this limit, a variety of so called “beam-beam effects” have usually led to intolerable operational conditions due to beam losses, beam size blowups, etc. From Eqs.(3) and (5) one can see that  the path to higher luminosity via higher beam intensity and smaller beam size almost automatically leads to higher beam-beam parameters. To get around the above-mentioned “beam-beam limit” several methods have been implemented over the decades, including a) making a careful choice of working tunes away from the most detrimental resonances; b) operation with very flat bunches (wide in horizontal plane and narrow in the vertical – see Eq.(12)); c) breaking beams into a significant number of bunches separated in all but a few interaction points, so the single bunch intensity $N$ is reduced; and, more recently, e) compensation of the beam-beam effects using electron lenses [9]; f) reduction of the strength of the beam-beam resonance in the “round beams” scheme with strongly coupled vertical and horizontal motion [10] and by g) using so called “crab-waist” collision method that beneficially modifies the geometry of the colliding bunch profiles only at the interaction points [11]. Despite all the inventions, the beam-beam effects are still considered to be setting a not fully resolved limit on the performance of most colliders.

There are two ways to achieve high luminosity within the beam-beam limit – either to increase the beam current $I=e\ f\ N$ or to decrease the beam emittance $\varepsilon$. In addition to the synchrotron radiation power which grows linearly with $I$, the major limitations on the current come from so-called coherent beam instabilities and from the demands of the radiation protection from inevitable particle losses. The instabilities are caused by beam interaction with the electromagnetic fields induced by the beam itself in the vacuum cambers and RF cavities, or caused by unstable clouds of secondary particles, like electrons or ions, which are formed around the circulating beams [12, 13]. Beam-based transverse and longitudinal feedback systems and electron/ion clearing (either by weak magnetic or electric field or by modulation of the primary beam current profile make the secondaries unstable) are now in routine use to avoid the coherent instabilities. Incoherent particle losses may have a variety of causes: unstable single particle motion, diffusion due to scattering on residual vacuum molecules or on the other particles within the bunch, high-frequency noises in the guiding magnetic field or the EM fields of RF cavities, ground motion and man-made vibrations, etc. To avoid damage or excessive irradiation of the accelerator components so they can be accessed and maintained if necessary in the tunnel, sophisticated collimation systems are utilized which usually employ a series of targets (which scatter the halo particles) and absorbers (which intercept the particles in dedicated locations) [14], or use more sophisticated techniques like collimation by bent crystals [15] or by hollow electron beams [16].

As for the quest for smaller beam emittances (phase space area) the paths of the lepton and hadron colliders were different. In the electron machines where any transverse errors are naturally damped by the synchrotron radiation, the emittance is determined by quantum fluctuations of the radiation which excite mostly horizontal oscillations – therefore, the challenge is to design focusing beam optics to minimize the effect in the horizontal plane, and avoid its coupling to the vertical plane, ideally below 1% [17]. Hadron collides cannot enjoy fast damping due to the synchrotron radiation, at least for energies less than 10 TeV, so for them the only ways to progress for were either generate low-emittance (high brightness) beams in the sources or arrange beam "cooling" (phase space reduction, usually at the low or medium energy accelerators in the injector chain), using either "stochastic cooling" or "electron cooling" methods, and the latter has been recently exemplified in the relativistic regime [18].

Operation of the colliders with progressively smaller and smaller beams brought up many issues relevant to alignment of magnets, vibrations and long-term tunnel stability [19]. Radiation backgrounds in high-energy physics detectors necessitated careful design of the accelerator-detector interface in high luminosity colliders. High energy physics demands for polarized beam collisions and very precise c.m. energy calibration of about $(\delta E/E)\sim 10^{-5}$ have been largely satisfied by, correspondingly, the development of polarized particle sources married with sophisticated methods to maintain beam polarization along the acceleration chain [20] and by novel method of "resonant depolarization" [21, 17].

**Table 1**: Past, present and possible future colliders; hadron colliders are in **bold**, lepton colliders in *Italic*, facilities under construction or in decisive design and planning stage are listed in parenthesis (…)

| | early 1990's | early 2010's | 2030's |
|---|---|---|---|
| Europe | **H*E*RA,** (**LHC**) | **LHC** (*Super-B*, **HL-LHC,** | **HE-LHC** |
| | *LEP (Dafne)* | **LH*e*C,** *E***NC**) | *CLIC?* |
| Russia | *VEPP-2, VEPP-4* | *VEPP-2000, VEPP-4M* | **NICA** ? |
| | (**UNK**, *VLEPP*) | (**NICA**, *Tau-Charm*) | *Higgs Factory* ? |
| United States | **Tevatron**, (**SSC**) | **RHIC** | *Muon Collider* ? |
| | *SLC, CESR*, (*PEP-II*) | (*e***RHIC**, *EL***IC**) | *PWLA/DLA* ? |
| Asia | *Tristan, BEPC* | *BEPC* | *ILC* ? |
| | (*KEK-B*) | (*Super-KEKB*) | *Higgs Factory* ? |
| Total | 9 (7) | 5 (9) | 1 + ? |

To conclude this chapter, we may say that colliders have had 50 glorious past years as not only many important particle discoveries were made at them, but they also initiated a wide range of innovation in accelerator physics and technology which resulted in 100-fold increase in energy (for each hadron and lepton colliding facilities) and $10^4$-$10^6$ fold increase of the luminosity. At the same time, it is obvious that the progress in the maximum c.m. energy has drastically slowed down since the early 1990's (and lepton colliders even went backwards in energy) – see Fig.2. Moreover, the number of the facilities in operation has dropped from 9 to 5, as indicated in Table 1 which lists all operational colliders as of the early 1990's and now (early 2010's) and accounts for the projects under construction or under serious consideration at this time (in parenthesis). Tectonic changes in the field happened in early 1990's when two flagship projects were terminated – 6 TeV c.m. proton-proton complex UNK [22] in Protvino, Russia in 1991 and the 40 TeV c.m. proton-proton Superconducting Super Collider in Texas, USA in 1993 [23]. Our current landscape shows the end of the Tevatron era (the 26 years long ~2 TeV c.m. energy proton-antiproton Collider Run ended in September 2011) and is dominated by the LHC at CERN. The Tevatron, LEP and HERA established the Standard Model (SM) of particle physics. The next generation of colliders is expected to explore it at deeper levels and to eventually lead the exploration of the smallest dimensions beyond the current SM. In the following chapters we outline possible colliders for the next 20 years and take a look into the issues and options for the even more distant future.

## Chapter 2: Next 20 Years: Physics, Technologies and Machines

The future of the collider is ultimately driven by the demands of particle physics, but should stay within the limits of the available technologies and financial resources. All the projects currently under construction or at the design stage (see Table 1) satisfy those three requirements and, thus have good prospects of becoming operational and deliver results in the next 20 years. Schematically they can be categorized by the area of the promising physics as follows:

*Energy Frontier :* the LHC luminosity upgrade project HL-LHC [24] will employ novel SC magnet technology based on the $Nb_3Sn$ superconductors for tighter focusing at the interaction points and quintuple the performance of the energy frontier machine by mid-2020's to $5\cdot10^{34}$cm$^{-2}$s$^{-1}$ with luminosity leveling at 14 TeV c.m. energy in proton-proton collisions and will enable to obtain about 250 fb$^{-1}$ of the integrated luminosity per year with ultimate goal of 3000 fb$^{-1}$ for both ATLAS and CMS experiments.

*Low-energy hadron collisions:* investigation of the mixed phase of quark–gluon matter and polarization phenomena at relatively low hadron energies has recently become of significant interest for the high energy physics community, and it is the main goal of the Nuclotron-based Ion Collider fAcility (NICA) currently under construction at JINR (Dubna, Russia) [25]. NICA will allow for the study of ion-ion ($Au^{+79}$) and ion-proton collisions in the energy range of 1-4.5 GeV/amu with average luminosity of $10^{27}$ cm$^{-2}$ s$^{-1}$ and also polarized proton-proton (5-12.6 GeV) and deuteron-deuteron (2-5.8 GeV/amu) collisions – in that regime luminosities up to $10^{31}$ cm$^{-2}$ s$^{-1}$ are foreseen. The plans indicate start of operation and first physics results later in this decade.

*Electron-hadron collisions:* deep inelastic electron-nucleon scattering is in the focus of a new electron-hadron collider project, the LHeC [26], in which polarized electrons of 60 GeV to possibly 140 GeV collide with LHC protons of 7000 GeV with design luminosity of about $10^{33}$ $cm^{-2}s^{-1}$. This would exceed the integrated luminosity collected at the previous *ep* collider HERA at DESY by two orders of magnitude in a 20 times wider kinematic range in the momentum transfer $Q^2$. Similar approach of reusing an existing beam facility and adding an accelerator for another species is taken in two collider projects in the US – eRHIC at BNL [27] and Electron-Ion Collider (ELIC) at JLab [28]. The eRHIC design is based on one of the existing RHIC(Relativistic Heavy Ion Collider) hadron rings which can accelerate polarized nuclear beam to 100 GeV/nucleon and polarized protons upto 250 GeV for, and a new 20-30 GeV multi-pass energy-recovery linac (ERL) to accelerate polarized electrons; the luminosity varies from $10^{33}$ $cm^{-2}s^{-1}$ to $10^{34}$ $cm^{-2}s^{-1}$ depending on the energy and species. The ELIC proposal re-uses the CEBAF 3-7 GeV polarized electron accelerator and requires the construction of a 30 to 150 GeV storage ring for ions (*p, d,* $^{3}He$ and *Li*, and unpolarized light to medium ion species). The attainment of very high luminosities in the ELIC, from $5{\cdot}10^{33}$ $cm^{-2}s^{-1}$ to $10^{35}$ $cm^{-2}s^{-1}$, an ERL-based continuous electron cooling facility is anticipated to provide low emittance and simultaneously very short ion bunches. Though with lower c.m. energy than LHeC, both of the projects in the US have the advantage of colliding both electron and ion species with polarized spins. It is believed that not more than one of the two can be supported and constructed.

Complementary physics programs can be realized at the proposed electron-nucleon collider ENC at the upcoming Facility for Antiproton and Ion Research FAIR at GSI Darmstadt (Germany) by utilizing the 15 GeV antiproton high-energy storage ring HESR for polarized proton and deuteron beams, with the addition of a 3.3 GeV storage ring for polarized electrons [29]. This will enable electron-nucleon collisions up to a center-of-mass energy upto 14 GeV with peak luminosities in the range of $10^{32}$ to $10^{33}cm^{-2}s^{-1}$.

*Electron-positron factories:* In the late 1990's – early 2000's, two asymmetric-energy $e{+}e{-}$ *B*-factories, the KEKB collider for the Belle experiment at KEK and the PEPII collider for the BaBar experiment at SLAC, had achieved tremendous success in confirmation of the Standard Model (SM) in the quark flavor sector and indicated that the Kobayashi-Maskawa mechanism is the dominant source of the observed *CP* violation in nature. Despite that, two fundamental questions remain unanswered in the flavor sector of quarks and leptons: a) it is not clear why the SM includes too many parameters and b) there is still a serious problem with the matter-antimatter asymmetry in the universe. To extend physics reach beyond two *B*-factories, much higher (by a factor of 40 or so) luminosity Super-*B* factories are either set up or considered for construction – one in Italy [30] and another in Japan [31]. Both are asymmetric-energy $e+e-$ colliders with beam energies of about 4 GeV and 7 GeV and with a design luminosity approaching $10^{36}$ cm$^{-2}$ s$^{-1}$, which is to be achieved via somewhat higher beam currents and very small beta-functions at the interaction points $\beta_y^* \sim 0.3$mm made possible by employment of the above mentioned "crab waist" scheme. The physics run of the Super-KEKB in Japan is expected in 2015 and the physics run will start in 2015. Ultimately, Belle II detector should collect 40 times more *B*-meson samples per second than its predecessor – roughly 800 *BB* pairs per second and accumulate an integrated luminosity of 50 ab$^{-1}$=50,000 fb$^{-1}$ by 2021.

Many similar technical solutions, e.g. the "crab waist", will also be employed in the project of TauCharm factory in Novosibirsk (Russia) [32] which calls for c.m. collision energy variable from 3 GeV to 4.5 GeV (from *J/psi* resonances to charm barions), luminosity in excess of $10^{35}$ cm$^{-2}$ s$^{-1}$ and longitudinal polarization of at least one (electron) beam.

If one will project at the very end of the next 20 years, then the landscape of the collider physics is much less certain, there are several directions to advance and the choice between the options will be based upon the results from the LHC. The relevant results are expected to be available starting in 2012-2013 (e.g., anticipated discovery of the Higgs boson) but they might easily slip well into the 2020's. Let us look into five possibilities for an after-LHC collider of 2030's.

*Higher energy LHC:* One of the most feasible opportunities is an energy upgrade of the LHC to 33 TeV c.m. proton-proton collisions [33]. The HE-LHC in the existing LHC tunnel will require 20T dipole magnets which are currently thought possible via combination of the NbTi, $Nb_3Sn$ and HTS (high-temperature superconductor) SC magnet technology. Such a collider could follow the HL-LHC and start operation in the early 2030's. Despite the (presumed) feasibility of the machine, its energy reach is limited to ~2.5 times the LHC energy and it is not fully clear yet whether such a (relatively) small energy advance will justify its construction.

*Higgs Factory:* The Higgs boson with mass $m_H$ of about 125 GeV has been recently discovered at the LHC, and the detailed studies and precise measurements of this unique spin-0 elementary particle might be of enough significant interest to justify construction of a $e+e-$ collider – a dedicated "Higgs factory". The maximum cross-section, and arguably the optimal centre-of-mass energy for studies of a number of Higgs boson properties, is at $E_{cm} \sim m_H + (110\pm10)$ GeV ~240 GeV, and several opportunities for the facility are now under discussion, including one based on the ILC-type linear collider (see below) as well as several ring-ring versions [34]. The biggest challenge for the latter is the requirement to replenish energy loss of electrons and positrons due to the synchrotron radiation of the order of 10 GeV per turn even in 20-km or longer tunnels - see Eq.(8) – that with necessity means extensive use of high gradient SC RF accelerating cavities. Other challenges toward attainment of the required luminosity of ~$10^{34}$ $cm^{-2}$ $s^{-1}$ (equivalent to 20,000 events per year under assumption of the $e+e- \rightarrow HZ$ cross section of about 200 femtobarn (fb)=$2\cdot10^{-34}$ $cm^2$) will be significant electric power consumption on the order of 100 MW needed for continuous acceleration of ~10 mA of beam current and the need for very small beam emittances and very large momentum acceptance of the ring to accommodate the energy losses at the interaction points (see discussion on the *beamstrahlung* effect below). A cost saving option of the Higgs factory in an existing tunnel, e.g., 26.7 km long LHC tunnel or 21 km long UNK tunnel, looks particularly attractive.

Alternative way for production of the Higgs bosons is in the reaction $\mu+\mu- \rightarrow H$ (so called *s-channel* reaction) which has advantageously large cross section for muons, $(m_\mu/m_e)^2$~40,000 times higher than for electrons, and (another advantage) needs a $\mu+\mu-$ collider with factor of two lower c.m. energy $E_{cm} \sim m_H$. The third advantage of that scheme is significantly smaller c.m. energy spread $\delta E_{cm}/E_{cm} \sim$ 0.01-0.003% (compared to ~0.2% for the $e+e-$ factories) that allows much better study of the outstandingly narrow width Higgs particle decays [35]. Production of ~4,000 events per year will require luminosity of at least $10^{31}$ $cm^{-2}$ $s^{-1}$ which seems to be very challenging because of the short muon lifetime and difficulties of the muon production (see discussion on high energy muon colliders below).

*Energy Frontier Lepton Collider:* It is presently widely believed that a multi-TeV lepton collider will be needed to follow the LHC discoveries. The physics program that could be pursued by a new lepton collider with sufficient luminosity, would include understanding the mechanism behind mass generation and electroweak symmetry breaking; searching for, and possibly discovering, super-symmetric particles; and hunting for signs of extra space-time dimensions and quantum gravity. By the beginning of the 2020's, the results obtained from the LHC will be expected to more precisely establish the desired lepton collider energy. The most viable options currently under consideration are $e+e-$ linear colliders ILC (International Linear Collider) [36] and CLIC (Compact Linear Collider) [37] or $\mu+\mu-$ Muon Collider [38]. Each of these options has its own advantages, challenges and issues [39, 37].

**Table 2**: Comparison of Lepton Collider alternatives

| | ILC | CLIC | MC |
|---|---|---|---|
| c.m energy, TeV | 0.5 | 3 | 1.5-4 |
| c.m. energy spread, rms | ~2% | >5% | ~0.1% |
| Luminosity , $cm^{-2}s^{-1}$ | $2{\cdot}10^{34}$ | $2{\cdot}10^{34}$ * | $(1\text{-}4){\cdot}10^{34}$ |
| Feasibility report | 2007 | 2012 | 2014-16 |
| Technical design | 2013 | 2016 | ~2020 |
| Number of elements | 38,000 | 260,000 | 10,000 |
| Hi-Tech length, km | 36 | ~60 | 14-20 |
| Wall plug power, MW | 230 | 580 | 120-200 |

* peak luminosity within 1% c.m. energy spread

The biggest challenge for the linear $e+e-$ colliders is to accelerate the particles to the design energy within a reasonable facility footprint and with as high as possible power conversion from the "wall-plug" to the beams. The ILC employs pulsed 1.3GHz SC RF cavities with average accelerating gradient of 33.5 MV/m, has the total length of the 0.5 TeV c.m. energy facility of about 31 km and has design power efficiency (beam power/total AC power) of about 8%. CLIC scheme is based on two-beam acceleration in 12 GHz normal conducting RF structures with average gradient of 100 MV/m, the total length of the main tunnel of 3 TeV c.m. collider is 48 km and overall beam power efficiency is ~5%. Both projects have in principle demonstrated technical feasibility of their key acceleration technologies. Both have very tight requirements on the beam emittance generated in several km long injection rings, emittance preservation in the main linacs where beam is subject of minuscule transverse kicks due to vibrations and other dynamic misalignments, and need for ultimate precision beam position monitors to stabilize beam trajectories on every shot using fast beam-based feedback systems. The stability tolerances are even tighter for the elements of several-km long "final focus" systems – accelerator beamlines to focus beams to unprecedented beam sizes of $\sigma^*_y/\sigma^*_x$=6nm/640nm in the ILC and 0.9nm/45nm in CLIC. Another "not-so-easy" to get around challenge is the c.m. energy spread induced by beamstrahlung (the energy loss caused by radiation of gamma quanta by the incoming electron due to its interaction with the EM field electron (positron) bunch moving in the opposite direction) during the very moment of collision of short bunches with rms length of $\sigma_z$ =50-300 μm, that for parameters of interest can be approximated as :

$$\frac{\delta E}{E} \propto \frac{\gamma N^2 r_e^3}{\sigma_x^2 \sigma_z} \quad , \tag{13}$$

and reach several % or even 10% - see Table 2. The induced radiation generates undesirable background in the detectors, makes handling of the beams after the collision more sophisticated and, most importantly, sets limitation on the energy resolution of the narrow resonances such as in the Higgs- and expected *Z'*-boson decay reactions.

Muons, which can be thought of as a heavy electrons, are essentially free of all synchrotron radiation related effects, which are proportional to the fourth power of the Lorentz factor $\gamma^4$, and, thus, $(m_\mu/m_e)^4 = (207)^4 = 2\times10^9$ times smaller. So, a multi-TeV $\mu^+\mu^-$ collider [39] can be circular and therefore have a compact geometry that will fit on existing accelerator sites (see Fig.5 for a possible schematic of a MC facility on the 6×7 km FNAL site ). The collider has a potentially higher energy reach than linear *e+e-* colliders, its c.m. energy spread in a 1.5-4 TeV $\mu^+$ $\mu^-$ collider can be as small as 0.1%, requires less AC power and operates with significantly smaller number of elements requiring high reliability and individual control for effective operation - see Table 2. Additional attraction of a Muon Collider(MC) is its possible synergy with the Neutrino Factory concept [40] as beam generation and injection complex of that facility and of a MC are similar (perhaps identical) [41]. As mentioned above, due to higher mass of the muon and superb energy resolution, a Higgs factory based on low(er) energy $\mu^+$ $\mu^-$ collisions is very attractive, too.

The biggest challenges of a MC come from the very short lifetime of the muon - $\tau_0$=2μs is just long enough to allow acceleration to high energy before the muon decays into an electron, a muon-type neutrino and an electron-type antineutrino ( $\mu^- \to e^- \nu_\mu \bar{\nu}_e$ ) – and from the methods of the muon production as tertiary particles in the reactions $pN \to \pi + ... \to \mu + ...$, so, the beams of muons are generated with very large emittances. A high-energy, 1-5 TeV c.m., high-luminosity $O(10^{34})$ cm$^{-2}$s$^{-1}$ muon collider seems will require a factor of $O(10^6)$ reduction of the 6-dimensional muon beam phase space volume (muon cooling). Though there has been significant progress over the past decade in developing the concepts and technologies needed to produce, capture and accelerate muon beams with high intensities on the order of $O(10^{21})$ muons/year, the feasibility of the high luminosity multi-TeV muon collider can be claimed only after demonstration of the fast ionization cooling of muons and resolution of the related issue of normalconducting RF cavities breakdown in strong magnetic fields. The latter is expected to be addressed by 2014-16, while convincing demonstration of the 6D cooling might take another 4 to 6 years [39].

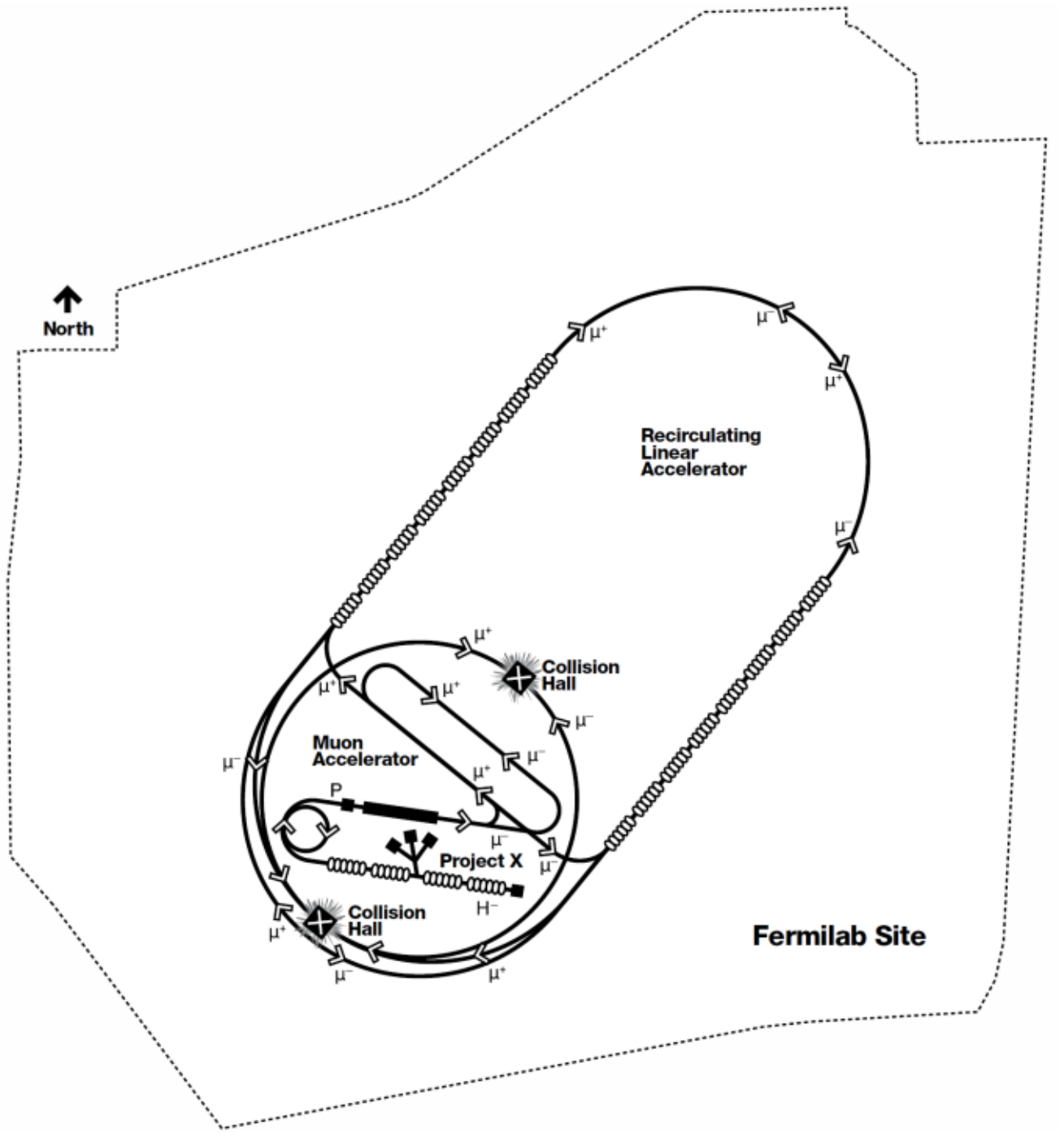

Figure 5: Schematics for a 4 TeV Muon Collider on FNAL site.

## Chapter 3: Beyond 2030's: New Mrethods and Paradigm Shift

Our forecast on the future of colliders beyond the 2030's will require several assumptions: on the available resources, on the desired science reach, and on the possible ways to the goal. Let us start with the money.

As of today, the world's particle physics research budget is some 3B$. That is about 1.2% of the world's total spending on basic science research (about 250B$). For comparison, global R&D funding of some 1,400B$ is about 2% of the world's GDP of ~70,000B$ (all numbers from [42]). We argue that the era of the exponential expansion for most out of ~200 sciences, see, e.g. [43] is gone, and for the particle physics, the new era of much more modest growth or even slow down of the financial support began in the 1990's - as one can also conclude from Fig.2 above. Despite the fact that most of today's scientific leaders in the scientific societies, governments and universities came of age during previous "golden era" and have not yet fully adjusted to new realities, we will base our predictions on the assumption that the field will stay approximately within the current financial limits, while still vigorously exploring new methods and directions. This immediately cuts a number of unrealistic possibilities, such as the Enrico Fermi ultimate accelerator or "globaltron" [44] – see Fig.6 – as its cost will exceed 20,000B$ even under modest estimate of 0.5B$ per kilometer of a high-tech accelerator.

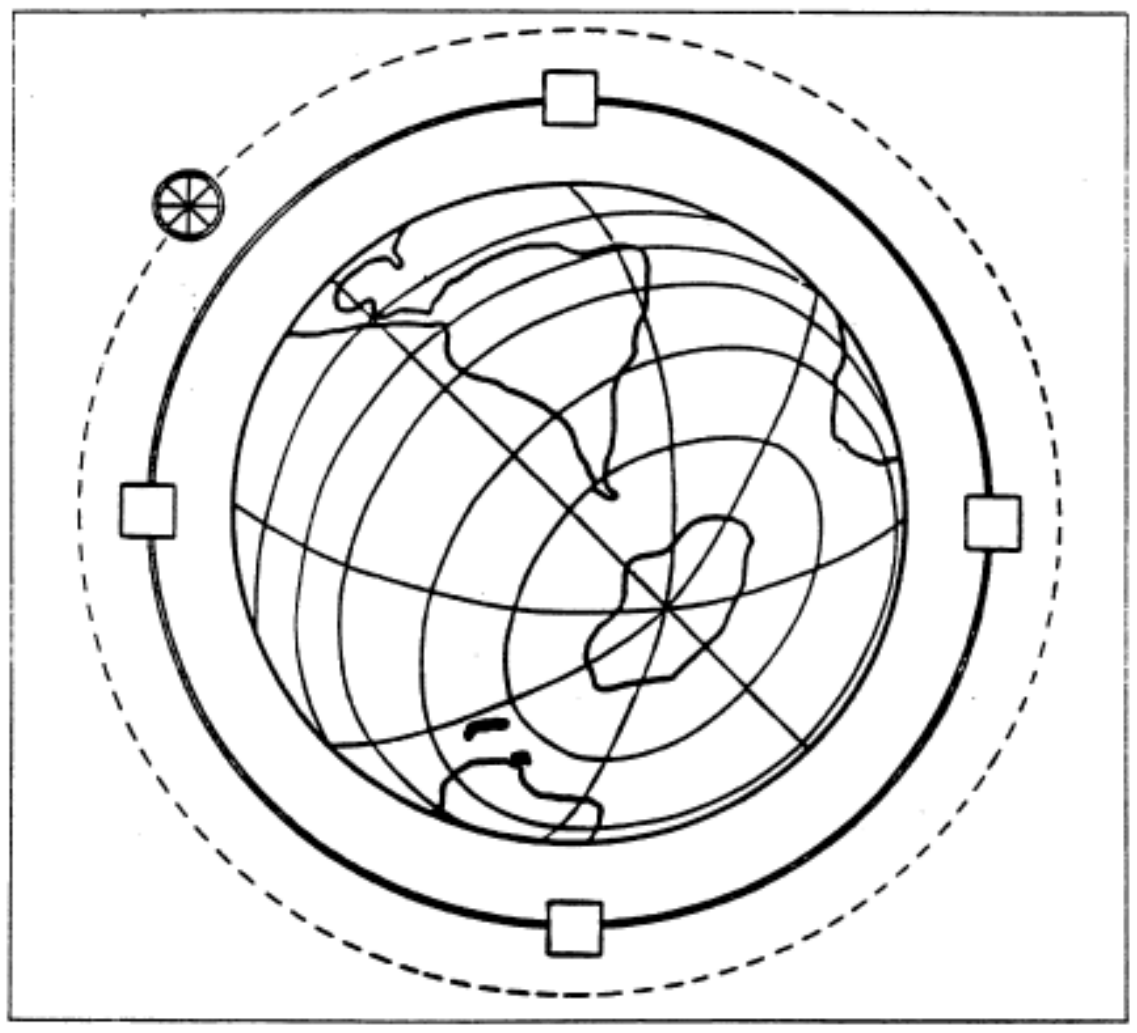

From a 1954 Slide by Enrico Fermi, University of Chicago Special Collections.

Figure 6: E.Fermi's "ultimate accelerator" circling the Earth.

The colliding beam facilities discussed in the preceding Chapter II, can be distributed in several logarithmically broad categories, depending on the cost of construction:

| | |
|---|---|
| Category I (under ~0.3B$) | NICA, ENC |
| Category II (0.3B$ - 1B$) | Super-*B* factories, Tau-Charm factory, eRHIC, ELIC |
| Category III (1B$ - 3B$) | Higgs factory, HL-LHC |
| Category IV (3B$-10B$) | HE-LHC, LHeC, Muon Collider, Higgs factory-ILC |
| Category V (10B$ - 30B$) | ILC, CLIC |

From what we know now, the Category I and II colliders can be built by one country, with relatively modest international contribution; Category III – requires significant cooperation at least within one of the big regions (Europe, Asia, America); Category IV machines will arguably need close cooperation of two or all three regions; while the Category V colliders must be truly international. Following our basic assumption (that the particle physics budget will not grow much beyond ~3B$ in current prices), it is hard to imagine that any of the projects in Category V are possible, as none of the possible scenarios – "do nothing but construction for 3-10 years" or "carve some 1/3 of the budget for construction and finish it in 10-30 years", or doing that at even slower pace – none of these seem realistic, especially for the energy reach being only comparable to the LHC. Category III machines appear totally feasible financially, but they only add to exploration of the phenomena discovered by the LHC, thus, there always will be an alternative "why don't we just run LHC longer?" Finally, we are left with Category IV colliders which appear to be on the border of financial affordability. There is a problem, though, with them – they have limited energy reach, equivalent to some 5 TeV c.m. energy in parton collisions (that is the Muon Collider energy, and the HE-LHC c.m. energy divided by an equivalent number of partons in proton 33/(6-10)=3-5 TeV). It is not whatsoever certain whether there be a demanding physics case for such kind of c.m. energies to justify the required significant investment.

Below we will offer some possibilities for physics reach of far future colliders within approximate budget of 10B$ (in current prices). The cost models of the modern colliders are quite complicated and do not fit the scope of this article, but one may safely assume that the future facility of such cost should not exceed 10 or few 10's of km in length and simultaneously require less than 10 to few 10's of MW of beam power (or correspondingly stay under ~100MW of AC wall power consumption). Can such colliders reach energies orders of magnitude beyond current – namely, 100-1000 TeV?

To get to the energies of interest within given footprint, one needs fast acceleration. At present, three main opportunities for high energy colliders are being actively discussed which can be schematically classified by the type of media used for acceleration – solid state structures, plasma and crystals. (Below we reference the most recent reviews.)

*Acceleration in dielectric structures:* Most of present day accelerators employ radiofrequency fields ($f_{RF}$ <10-30 GHz) in resonant normal-conducting or SC structures powered by conventional RF sources and are in general limited to gradients of ~100 MV/m due to surface breakdown phenomena. Combination of direct beam excitation (or wakefields radiated by a short intense ''driving'' bunch of electrons propagating in a high impedance environment) and hard dielectric materials (quartz, diamond, garnets, etc – which are characterized by lower power losses and higher breakdown gradients than metals) for structure fabrication, allow accelerating gradients of ~100 MV/m and ~1 GV/m in microwave $O(10)$ GHz and THz dielectric structures [45], correspondingly. Conceptually, a dielectric wakefield accelerator (DWA)-based linear collider would consist of a large number of ~100 GeV modules (stages) with some ~0.3 GeV/m gradient each driven by a separate ~1 GeV high intensity electron beam. Even without going into difficulties associated with staging, cost and power considerations, it is hard to imagine that more than 3 TeV c.m. energy DWA facility can fit within a 10 km site.

Further increase of the gradient to ~1-3 GV/m is thought to be possible in μm scale dielectric structures driven by lasers operating in optical or near-infrared regime [46]. In various options either external fiber lasers are coupled to the structures or semiconductor lasers can be integrated on the same slab right next to the microcells they power. Advantage of such an approach is that laser power sources can operate at very high repetition rates of ~10-100 MHz, that helps to get higher luminosity – see again Eq.(5) – but again, a 10 TeV c.m. energy collider will require ~12 km of the total linac length [47]. To be of note that staging of sequential accelerating modules made of smaller size structures (microns vs millimeters) will require proportionally tighter synchronization, alignment and mechanical stability tolerances to keep beam trajectories and emittances under control.

*Acceleration in plasma:* In the past decade plasma-wakefield acceleration (PWA) methods has become of great interest because of the promise to sustain extremely large acceleration gradients. Electric fields due to charge separation in dense plasma are of the order of

$$E_0 = \frac{m_e c \omega_p}{e} \approx 100[\frac{GeV}{m}] \cdot \sqrt{n_0[10^{18} cm^{-3}]} \quad , \tag{14}$$

where $\omega_p=(4\pi n_0 e^2/m_e)^{1/2}=2\pi c/\lambda_p$ is the electron plasma frequency, $n_0$ is the ambient electron number density, $\lambda_p \approx 30[\mu m]\cdot(n_0\ [10^{18} cm^{-3}])^{1/2}$ is the plasma wavelength which sets the characteristic scale length of the wakefield. For example, generation of the PWA gradients on the order of 30-100 GV/m at plasma densities of $n_0=10^{17}$-$10^{18}$ cm$^{-3}$ have been already demonstrated in small scale (few cm to a meter) experiments [47]. There are two ways to separate electrons and ions in the plasma – by lasers and by external beams. In the laser–plasma accelerators, a longitudinal accelerating electric field is generated by the ponderomotive force of an ultraintense laser pulse with duration on the order of the plasma period. This force, proportional to the gradient of the laser intensity, pushes the plasma electrons out of the laser beam path, separating them from the less mobile ions. This creates a travelling longitudinal electric field, in the wake of the laser pulse, with a phase velocity close to the speed of light – i.e., as needed for accelerating relativistic particles. The structure of the wakefield has broad phase space where negatively charged particles can be both accelerated and focused. Depletion of the laser energy determines the energy gain and the length of a single acceleration stage after which a new laser pulse must be coupled into plasma for further acceleration. Practical considerations indicate that this stage-to-stage distance is on the order of 1 m, and minimization of the total linac length requires operation at relatively low densities $n_0$~$10^{17}$cm$^{-3}$ and energy gain per stage of 10 GeV and average gradient ~5TeV/km [48].

Alternative LWA concept involves the passage of an ultra relativistic electron bunch through a stationary plasma (either pre-formed by ionizing a gas by a laser or through field-ionization by the Coulomb field of the relativistic electron bunch itself). The plasma electrons are repelled by the bunch. Generating large amplitude wakefields requires short high density electron bunches compressed in all three spatial dimensions to sizes smaller than $\lambda_p$ – e.g., bunches of few $10^{10}$ electrons of few μm's spot size and ~10 μm long. The wake produces a high gradient accelerating field Eq.(14) and transverse focusing for a negatively charged witness bunch behind the drive bunch [49]. The figure of merit in wakefield accelerators (plasma or dielectric) is the transformer ratio $R_T$=(maximum accelerating field behind the drive bunch)/ (the maximum decelerating field inside the drive bunch) which is limited to ~2 for a finite length, longitudinally symmetric drive bunches and can reach somewhat reach higher values only for specially prepared triangular bunch current shapes. A proposed beam-PWA collider design [50] consists of a conventional 25 GeV electron drive beam accelerator, which produces drive bunches distributed in counter-propagating directions to many (dozens to hundreds, depending on the final energy) one meter long plasma cells for both the electron and positron arms of the collider. Though each cell provides 25 GeV energy gain ($R_T \approx 1$), an average geometric accelerating gradient reaches only 0.25 TeV/km after taking into account some 100 meter long cell-to-cell gaps needed to bring up a fresh drive beam to the next cell from a single source of the high intensity 25 GeV electron bunches. As in other schemes, serious issues associated with staging and transfers from one cell to another (alignment, synchronization, etc) are anticipated. The number of stages can be significantly reduced if higher energy drive beams are available, e.g., some 0.6 TeV energy gain might be possible in a single ~400m plasma cell driven by 1 TeV protons if such high energy proton bunches can be compressed longitudinally to under 1mm bunch length and kept tightly focused transversely [51]. Without going into practical considerations and just projecting such gradients further one can think of an ultimate ~10 TeV collider within 10 km footprint.

_Acceleration in crystal channels:_ The density of charge carriers (conduction electrons) in solids $n_0$~$10^{22\text{-}23}$ $cm^{-3}$ is significantly higher than what was considered above in plasma, and correspondingly, the longitudinal fields of upto 100 GeV/cm or 10 TV/m are possible – see Eq.(14). The new effects at higher densities are due to intense energy radiation in high fields and increased scattering rates which result in fast pitch-angle diffusion over distances of $l_d$ *~1[m]·E[TeV]*. The latter leads to particles escaping from the driving field; thus, it was suggested that particles are accelerated in solids along major crystallographic directions, which provide a channeling effect in combination with low emittance determined by an Ångström-scale aperture of the atomic "tubes." Channeling with nanotubes is also being discussed. Positively charged particles are channeled more robustly, as they are repelled from ions and thus experience weaker scattering. Radiation emission due to betatron oscillations between the atomic planes is thought to be the major source of energy dissipation, and the maximum beam energies are limited to about 0.3 TeV for positrons, $10^4$ TeV for muons and $10^6$ TeV for protons [52]. X-ray lasers can efficiently excite solid plasma and accelerate particles inside a crystal channel waveguide, though ultimate acceleration gradients ~10TeV/m might require relativistic intensities, exceeding those conceivable for x-rays as of today [53]. Moreover, only disposable crystal accelerators, e.g., in the form of fibers or films, are possible at such high externally excited fields which would exceed the ionization thresholds and destroy the periodic atomic structure of the crystal (so acceleration will take place only in a short time before full dissociation of the lattice). For the laser and plasma fields of about 1 GV/cm=0.1 TV/m or less, reusable crystal accelerators can probably be built which can survive multiple pulses [54]. Side injection of powerful x-ray pulses into continuous fiber of 0.1 – 10 km long fiber allows to avoid multiple staging issues intrinsic to other methods and reach 10-1000 TeV energies, if the imperfections like crystal dislocations are kept under control and unintended crystal curvatures are less than inverse "critical" radius $R_c$ *~2[m]·E[TeV]* – so the channeling conditions remain [15].

**Table 3**: Far-reach particle colliders

| | Dielectric based | Plasma based | Crystal channeling |
|---|---|---|---|
| Accelerating media | microstructures | ionized plasma | solid crystals |
| Energy source: option 1<br>option 2 | optical laser<br>$e^-$ bunch | $e^-$ bunch<br>optical laser | x-ray laser |
| Preferred particles | any stable | $e^-$, $\mu^-$ | $\mu^+$, $p^+$ |
| Max accelerating gradient | 1-3 GV/m | 30-100 GV/m | 0.1-10 TV/m |
| c.m. energy reach in 10 km | 3-10 TeV | 3-50 TeV | $10^3$-$10^5$ TeV |
| # stages/10 km: option 1<br>option 2 | $10^5$-$10^6$<br>$10^4$-$10^5$ | ~100<br>$10^3$-$10^4$ | ~1 |

It is to be noted that due to the imposed facility footprint limit, limited bending fields available and troubles with synchrotron radiation losses, the circular colliders do not seem conceivable for ultra-high c.m. energies ~10-100 times the LHC and one with necessity comes to a linear configuration - as in Fig.1c. Out of the purely energy gain arguments, heavier particles are preferred in all novel acceleration methods because they radiate less - once again, we refer to the synchrotron radiation energy loss formula Eq.(8) – and one might argue that acceleration of electrons and positrons beyond ~1-3 TeV is impractical. Also, even at that energy, the c.m. energy spread due to beamstrahlung of electrons at the interaction point Eq.(13) becomes prohibitively large at any other practical beam parameters and special measures will need to be taken, e.g., conversion of electrons and positrons in high energy $\gamma$-quanta and $\gamma$-$\gamma$ collisions [55]. Out of the remaining options of heavier particles, protons seems to be the only choice for c.m. energies beyond $10^4$ TeV=10 PeV; while in the context of the next 2-3 decades, when we do not envision energies of practical realization greater than 1 PeV, muons are much more attractive option because i) they are point-like particles und, contrary to protons, do not carry an intrinsic energy spread of elementary constituents; and ii) they do not have issues associated with nuclear interactions with accelerating media of plasma or solids. The fact that muons are not stable particles and decay as $dN/dt=-N/\gamma\tau_0$ becomes irrelevant in the fast acceleration schemes as the survived beam fraction at the final energy $E$ :

$$\frac{N}{N_0} \approx \left(\frac{m_\mu c^2}{E}\right)^\kappa \quad , \tag{15}$$

is very close to 1 as soon as the exponent $\kappa=(m_\mu c/\tau_0 G)<<1/\ln(E/m_\mu c^2)$ or, conversely, the average accelerating gradient $G$>>3 MeV/m – the condition that easily holds for any scheme considered above (see Table 3). Note that at $G$>>0.3 TeV/m, one can think of accelerating tau leptons, though production of needed quantities of $\tau$ particles is questionable.

Following Ref.[54], let us explore possible luminosity reach of ultimate energy colliders with $E_{\rm cm}$>100 TeV. Given that acceleration of heavy particles in solid media/crystals is the technology of choice, it seems to be reasonable to limit the minimal overlap area of the colliding beams to the crystal lattice cell size $A$~1 Å$^2$=10$^{-16}$cm$^{-2}$ and to assume that the crystals of each collider arm will be aligned channel to channel. The other factor in the luminosity formula (3) is the number of particles $N$ which can not be made arbitrary high due to the beam loading effect. Effective acceleration with transfer ratio $R_T$~1 is possible only if the number of particles in the beam does not exceed the number of particles in the plasma volume excited by external source. Such volume is about 100 $\lambda_p$ (longitudinal extent before excitation decays) × $\lambda_p^2$ that results in $N_0$~10$^3$ particles per individual bunch. Of course, exciting many parallel atomic channels $n_{ch}$ will proportionally increase the luminosity $L=f\cdot N^2/A=f\cdot 10^{16}\cdot 10^6\cdot n_{ch}$ [cm$^{-2}$s$^{-1}$] – which can reach 10$^{30}$cm$^{-2}$s$^{-1}$ at, e.g., $f$=10$^6$ Hz and $n_{ch}$~100. Exceeding the value of the product $fn_{ch}$ beyond 10$^8$ Hz can be very costly as the total beam power $P= fn_{ch}\cdot NE$ will exceed 16 MW per beam which we consider a practical limit from our collider cost considerations (see above). It might be beneficial instead to attempt to combine (focus) all the channeling beams into one using some kind of *crystal funnel* and, thus, gain a factor of $n_{ch}$ in the luminosity. Overall, in the power-limited scenario, the luminosity scales at very high energies as:

$$L[cm^{-2}s^{-1}] \approx 4\cdot 10^{33\div 35} \times \frac{P^2[MW]}{E^2[TeV]\cdot fn_{ch}[10^8 Hz]} \quad (16).$$

The performance of the ultimate energy colliders can not increase with energy – either it is independent of it (if total beam power is small) or falls as $\sim 1/E^2$. This fact, if not overturned by some future invention, indicates the need of the paradigm shift in the high energy particle physics because so far the performance goals of new energy frontier facilities scaled $L\sim E^2$, reflecting the fact that many important cross sections fall as $\sigma_{\rm int} \sim 1/E^2_{\rm cm}$. Seemingly, the physics reach of colliders of the future will be limited to either resonances (phenomena with unusually high production rates at certain energies, indicating new particles) or other high cross-section reactions. Still, even scaling as in Eq.(16) is much better than what other alternative sources of ultra-high energy particles can offer. For example, the number of cosmic ray events falls with the c.m. energy $E_{cm} \approx (2Emc^2)^{1/2}$ even faster, approximately as $\sim 1/E^7_{cm}$ and at the highest energies of $E_{cm}$~500 TeV, or $E$~$10^8$ TeV, the event rates are extremely low - of about few per week even for the largest area observatories.

## Conclusions:

The colliding beam method was a smashing success so far - almost 3 dozens colliders have been built over the past half-century and c.m. energies of about 10 TeV have been achieved. At the same time, the pace of the energy progress has greatly slowed down due to increasing size, complexity and cost of the facilities, and as the result, the number of colliders currently in operation is about half of what we had 20 years ago. The prospects of facilities for the next 20 years are not very clear today, they will be dependent on the discoveries at the current machines, foremost, at the LHC.

It seems that economic realities will impose severe constrains for any far-future collider beyond 2030 to be built under about 10B$ at current price, within footprint of some 10 km, and with total electric power consumption of 10's to 100MW. As discussed in the previous chapter, there are possibilities which even currently conceivable methods can offer to reach ultra high energies on the order of 100-1000 TeV within the abovementioned limits. The quest for the energy will come at the price of the expected luminosities and will require at least three paradigm shifts : 1) development of the new technology based on ultrahigh acceleration gradients ~0.1-10 TeV/m in crystals; 2) acceleration of heavier particles, preferably, muons; and 3) new approaches to physics research with luminosity limited to ~$10^{30\text{-}32}$ $cm^{-2}s^{-1}$.

As any other shift in the mainstream accelerator technology, the required switch to acceleration of muons in linear crystal structures will take a decade or two for an R&D program to address several key issues: a) development of economical high-intensity coherent X-ray sources, e.g., based on table-top ~GeV scale electron accelerators [56, 57]; b) understanding the most effective mechanisms of coupling the X-ray power to the excitation of the lattice - that can probably be studied even at the existing high power coherent X-ray sources, like LCLS in the US or Spring-8 in Japan; c) efficient production, injection and manipulation of nm-size muon beams (this program can effectively gain momentum out of the current research toward a muon collider); d) methods of combination of multiple crystal channeling beams into one (experiments at existing high energy proton machines might provide an important input there). As in the past, as soon as the main technology issues are address, a great boost to the new development can be given by a test facility where the new acceleration methods are used for exploration of interesting “low-energy” physics – e.g., a “table-top” factory of $\omega$, $\psi$, $\tau$ or even $Z$, $W$ or Higgs particles with decent luminosity and relatively low cost.

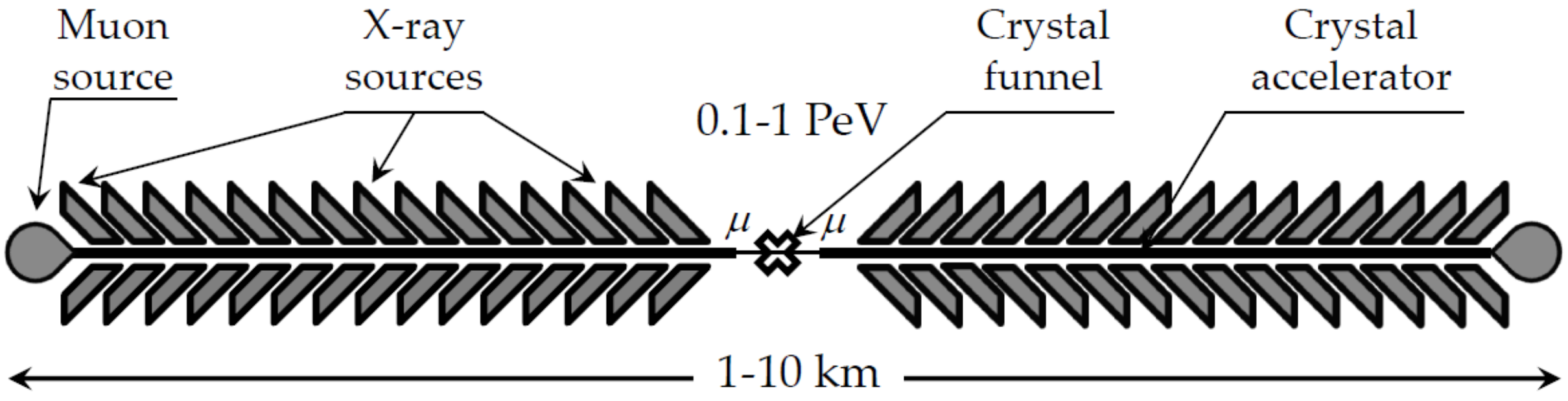

Figure 7: Concept of a linear X-ray crystal muon collider.

The author wishes to thank C.Hill, A.Valishev, E.Eichten and S.Henderson for helpful discussions.

**References**

1. Haussecker E F, Chao A W *Physics in Perspective* **13** 146 (2011)
2. Kerst D et al *Phys. Rev.* **102** 590 (1956)
3. O’Neil G K *Phys. Rev.* **102** 1418 (1956)
4. Tigner M, *Nuovo Cimento* **37** 1228 (1965)
5. Edwards D A, Edwards H T *Rev. Accel. Sci. Tech.* **1** 99 (2008)
6. Chao A W, Tigner M, eds. *Handbook of Accelerator Physics and Engineering* (World Scientific, 1999)
7. Skrinsky A in *Proc. 1995 IEEE Part. Accel. Conf. (Dallas)*, p.14;
   Skrinsky A *Sov. Phys. Usp.* **25** 639 (1982);

Budker G *Sov. Phys. Usp.* **9** 534 (1967)

8. Tollestrup A, Todesco E *Rev. Accel. Sci. Tech.* **1** 185 (2008)
9. Shiltsev V *Phys. Rev. Lett.* **99** 244801 (2007)
10. Danilov V, et al in *Proc. 1996 European Part. Accel. Conf. (Spain)*, p.1149
11. Raimondi P, Shatilov D, Zobov M, arxiv: physics/0702033 (2007)
12. Ohmi K *Phys. Rev. Lett.* 75 1526 (1995)
13. Raubenheimer T, Zimmermann F *Phys. Rev. E* **52** 5487 (1995)
14. Mokhov N, et al *JINST* **6** T08005 (2011)
15. Biryukov V, Chesnokov Yu, Kotov V *Crystal Channeling and Its Application at High Energy Accelerators* (Springer–Verlag, Berlin, Heidelburg, 1997)
16. Stancari G, et al *Phys. Rev. Lett.* **107** 084802 (2011)
17. Brandt D, et al *Rep. Prog. Phys.* **63** 939 (2000)
18. Nagaitsev S, et al *Phys. Rev. Lett.* **96** 044801 (2006)
19. Shiltsev V *Phys. Rev. ST Accel. Beams* **13** 094801 (2010)
20. Alexeev I, et al *Nucl. Instr. Meth. Phys. Res. A* **499** 392 (2003)
21. Shatunov Yu, Skrinsky A *Particle World* **1** 35 (1989)
22. Yarba V in *Proc. 1991 IEEE Part. Accel. Conf. (San Francisco)*, p.2913
23. Wojcicki S *Rev. Accel. Sci. Tech.* **2** 265 (2009)
24. Rossi L in *Proc. 2011 Intl. Part. Accel. Conf. (Spain)*, p.908
25. Trubnikov G, et al in *Proc. 2010 Russian Part. Accel. Conf. (Protvino)*, p.14
26. Klein M in *Proc. 2011 Intl. Part. Accel. Conf. (Spain)*, p.908
27. Ptitsyn V, et al in *Proc. 2011 Intl. Part. Accel. Conf. (Spain)*, p.3726
28. Ahmed S, et al in *Proc. 2011 IEEE Part. Accel. Conf. (New York)*, p.2306
29. Lehrach A, *et al J. Phys.: Conf. Ser.* **295** 012156 (2011)
30. Bagiani M, Raimondi P, Seeman J, arxiv:1009.6178
31. Abe T, et al arxiv:1011.0352
32. Levichev E *Phys. Particles and Nuclei Lett.* **5** 554 (2008)
33. *The High Energy Large Hadron Collider* Preprint CERN-2011-003 (2011)
34. Blondel A, Zimmermann F, arxiv:1112:2518
35. C.Ankenbrandt, et al *Phys. Rev. ST Accel Beams* **2** 081001 (1999).
36. *ILC Reference Design Report* ILC-Report-2007-001, http://www.linearcollider.org
37. Delahaye J P *Mod. Phys. Lett. A* **26** 2997 (2011)

38. S.Geer *Annu. Rev. Nucl. Part. Sci.* **59**:347–65 (2009)
39. Shiltsev V *Mod. Phys. Lett. A* **25** 567 (2010)
40. *The Neutrino Factory Int'l Scoping Study Accelerator Working Group Report*, *JINST* **4** P07001 (2009)
41. Poklonskiy A, Neuffer D *Int.J.Mod.Phys*. **A24** 959 (2009)
42. 2012 Global R&D Funding Forecast *Batelle R&D Magazine* (Dec. 2011)
43. Shiltsev V *Mod. Phys. Lett. A* **26** 761 (2011)
44. Cronin J (ed.) *Fermi Remembered* (University of Chicago Press, 2004)
45. Gai W in AIP *Conf. Proc.* **1086** 3 (2008)
46. Bermel P, et al *ICFA Beam Dynamics Newsletter* **56** 91 (2011)
47. Leemans W *ICFA Beam Dynamics Newsletter* **56** 10 (2011)
48. Schroeder C, et al *Phys. Rev. ST Accel. Beams* **13** 1013014 (2010)
49. Muggli P in *AIP Conf. Proc.* **1299** 52 (2010)
50. Seryi A *Nucl. Instr. Meth A* **623** 23 (2010)
51. Caldwell A, Lotov K, Pukhov A, Simon F *Nature Phys.* **5** 363 (2009)
52. Dodin I, Fisch N *Phys. Plasmas* **15** 103105 (2008)
53. Tajima T, Cavenago M *Phys. Rev. Lett.* **59** 1440 (1987)
54. Chen P, Noble R *AIP Conf. Proc.* **398** 273 (1997)
55. Telnov V arxiv:0908.3136
56. Schlenvoigt H, et al *Nature Phys.* **4** 130 (2008)
57. Kneip S, et al *Nature Phys.* **6** 980 (2010)